\documentclass[12pt,a4paper]{article}
\usepackage[latin1]{inputenc}
\usepackage{amsmath}
\usepackage{amsfonts}
\usepackage{amssymb}
\usepackage{graphicx}
\title{\bf Graphane - material for hydrogen storage, breathers and kinks.}
\author{Matej Hudak \\ {\it Matej Hudak's Lab A,} Stierova 23, SK - 040 23  Kosice, Slovak Republic 
	\and
	Ondrej Hudak \\ {\it Matej Hudak's Lab A,} Stierova 23, SK - 040 23  Kosice, Slovak Republic 
	\footnote{Corresponding author}\\
	{hudako@mail.pvt.sk}}
\date{June 18, 2020}

\begin{document}
	\maketitle{}
	
\newpage
	\vspace{1in}
	\begin{abstract}
In this paper we study the graphane. The Frenkel-Kontorova model on hexagonal lattice was used. We studied the case of one H atom above the C atom in the plane of graphane (we used the approximation of the hexagonal lattice in the plane). Continuous limit of the  Lagrange-Euler equations is found from the Hamiltonian for $H$ atoms motion, they enabled us to study kink and breather excitations of $H$ atoms in the $H$ plane above the $C$ plane. We have found that there are three cases in the one $H$ atom motion  The case $1$, when the $H$ atom is at the position which is below the position at which it is desorbed. Then the motion of this $H$ atom at time $t_{h}$ is described. The case $2$, when the $H$ atom is at the position of the suppressed atom $H$ in the direction to the $C$ (nearer) atom. This $H$ atom will be desorbed from the graphane going through the minimum of the potential energy and then through the point of desorption. Its motion of at time $t_{h}$ is described. The case $3$, when the $H$ atom is near the position of small oscillations near the potential energy minimum. The position of the atom $H$ at the time $t_{0}$ is the position to which the atom $H$ was excited with external force. The lattice of $H$ atoms in graphane may be excited as described by the kink solution of the Sine-Gordon equation. The kink has its velocity $U$, $ U^{2} < 1$, and in time $T$ and in $X^{'}$ coordinate direction localization. The Sine-Gordon equation has the breather solution in the $X^{'}$ direction. There
$\omega$ is the frequency of the breather, $T_{0}$ and $X^{'}_{0}$ are in time $T$ and in $X^{'}$ direction localization.  
	\end{abstract}

DOI: 10.13140/RG.2.2.24192.66565

\newpage
	
	\tableofcontents{}

\newpage
\section{Introduction.}

It is well known \cite{DKBV} that periodic discrete defect-containing systems, in addition to traveling waves, support vibrational defect-localized modes, if a system is nonlinear, it can support spatially localized vibrational modes in the absence of defects, discrete breathers (DBs). In \cite{BK} there are discussed discrete breathers (DB), their existence and stability, the discrete nonlinear  Schr\"{o}dinger equation, dark breathers and rotobreathers.  In general there are perturbed Sine-Gordon breathers, large-amplitude breathers and small-amplitude breathers, breathers collisions, many-soliton effects, fractal scattering, soliton cold gas, impurity modes, and soliton interactions with impurities. Discrete breathers are discussed in \cite{CMK}. A special emphasis is on the $\phi^{4}$ model. Spectral stability is studied numerically, as well as in special limits (such as the vicinity of the anti-continuum limit of vanishing coupling) analytically. 
In \cite{Sh} a family of solitary waves in Frenkel-Kontorova model and its continuum and 
quasi-continuum approximations are studied. Malomed \cite{BAM} studied nonlinearity and discreteness, and solitons in lattices. Homoclinic traveling waves in discrete Sine-Gordon model with nonlinear interactions on $2d$ lattices is studied in \cite{SMB} by means of critical point theory. The author obtains sufficient conditions for the existence of such solutions.
Dmitriev \cite{SVD} discusses discrete breathers in crystals. An overview of the dynamics of the Frenkel-Kontorova model, is presented in \cite{BK1}.The model describes the motion of a chain of interacting particles subjected to an external on-site periodic potential. Physically important generalizations of the model are discussed including non-sinusoidal on-site potentials and anharmonic potentials.

Carbon nanostructures are reviewed in \cite{STN}. After the discovery of graphene there was development of graphene derivatives. The $2d$ carbon allotropies of graphene are as graphyne and graphiyne. Two hydrogenated derivatives are graphone and graphane. Fluorinated graphene (fluorographene) is another structural derivative of graphene. The oxidized form of graphene (GO) can also be considered as a graphene derivative. The structure, properties, and applications of few significant graphene derivatives have been discussed in \cite{STN[48]} - \cite{STN[53]}. Graphane is the fully hydrogenated derivative of graphene, which is consisted of $sp^{3}$ carbon-carbon bonds only. Graphane has two conformations: chair-type and 
boat-type. The calculated carbon-carbon chain length in the chair-type conformer of graphane is similar to that of diamond and much higher than in graphene because of the $sp^{3}$ bonding characteristics. The binding energy of boat-type graphane is (calculated) higher than other hydrocarbons - like benzene, acetylene. The formation of graphane can be confirmed from the Raman spectroscopy. In comparison with the Raman spectrum of graphene, the $G$ and $2d$ bands broaden and additional $E$ band appears for graphane. The previous studies revealed that the band gap steadily increases with increasing the degree of hydrogenation. Graphane is much softer than graphene and exhibits lesser Poisson ratio. The fully hydrogenated graphane shows semiconducting nature. Hetero-atom - doped graphane is reported to exhibit semiconducting nature. Li-doped graphane exhibited high hydrogen storage capacity. Graphane shows promising application potentiality in the field of hydrogen storage, biosensing, and spintronics.

Graphane is a two-dimensional system consisting of a single layer of fully saturated ($sp^{3}$ hybridization) carbon atoms \cite{FALG}. Colossal carbon tubes \cite{FALG[1]} and graphene \cite{FALG[2]} were studied. Graphene was studied, see \cite{FALG[2]} - \cite{FALG[3]} and \cite{FALG[4]}. It has a two-dimensional structure of carbon atoms with $sp^{2}$ orbitals and interesting electronic and mechanical properties. Theoretically was predicted a related structure, graphane, in \cite{FALG[5]}. In \cite{FALG[5]} it was predicted that graphane has $C-H$ bonds in an alternating pattern. Two most stable conformations are the chair-like ($H$ atoms alternating on both sides of the plane) and boat-like ($H$ atoms alternating in pairs), see in \cite{FALG[5]} the Figure 1. A third member of these two-dimensional planar carbon structures is called graphyne \cite{FALG[6]} - \cite{FALG[8]}. Molecular fragments have been synthesized \cite{FALG[7]}. Graphane-like structures
experimental evidences have been reported in \cite{FALG[9]} and \cite{FALG[10]}. Elias et al. \cite{FALG[11]} demonstrated the existence of graphane formation from graphene membrane through its hydrogenation. They demonstrated that this process is reversible. Use of graphene is in the graphene - based devices (the electronic gap values in graphanes is controlled by the degree of hydrogenation) \cite{FALG[11]} and \cite{FALG[12]}. The $H$ incorporation results in altering the $C$ $sp^{2}$ hybridizations to $sp^{3}$ ones. Note, that the experiments over $SiO_{2}$ substrates (only one membrane side exposed to $H+$ ) produced a material with different properties. Detailed studies of hydrogen atoms on graphene have been reported in \cite{FALG[13]} -   \cite{FALG[20]}.

Electrical properties of large-area graphene were studied in \cite{BMW} enabled by fabricating large-area graphene films for transparent electrodes, barriers, electronics, telecommunication and other applications. In $2006$ the price of commercially available single layer graphene was 1 euro per $\mu m^{2}$. A decade after in $2016$, the price is approaching $1$ euro per $cm^{2}$. It is noted that Kobayashi and colleagues demonstrated preparation of $100$ meter long continuous graphene sheets transferred to a PET polymer substrate. The paper overview also other non-contact mapping techniques that provides information on the electrical properties: eddy current testing, scanning microwave impedance and Raman spectroscopy.

In \cite{KBLLDZ} clusters of discrete breathers in graphane were studied by molecular dynamics simulations. Graphene has high specific surface area, low weight and high strength. It is chemically inert. The paper \cite{KBLLDZ} discusses hypothesis that gap DBs (gap discrete breathers) can play an important role in the dehydrogenation of graphane. Existence of DBs was shown in $2d$ materials as graphene \cite{KBLLDZ[19]} and \cite{KBLLDZ[20]}, on the edge of graphene nanoribbons \cite{KBLLDZ[21]} and \cite{KBLLDZ[22]}, in boron nitride \cite{KBLLDZ[23]}, and in other carbon polymorphs \cite{KBLLDZ[24]} and \cite{KBLLDZ[25]}. In \cite{KBLLDZ} clusters of DBs consisting of two and three closely placed DBs are studied. Energy exchange between DBs is observed and discussed in relation to graphane dehydrogenation. Effect of temperature on the life time of DB is shown there for $50^{o}$ $K$ for single DB in thermal equilibrium. As is noted \cite{KBLLDZ} hydrogen is an ideal energy fuel. The physisorption of hydrogen on materials with a high specific surface area is possible. The hydrogenation of carbon nanotubes (CNTs) is discussed in \cite{KBLLDZ[1]} - \cite{KBLLDZ[3]}. The difference between CNTs and high surface area graphite is the curvature of the graphene sheets and the cavity inside the tube. The investigation \cite{KBLLDZ[2]} of hydrogen adsorption inside CNTs has shown that it is energetically more favorable for hydrogen atoms to recombine and form molecules. Different
nanostructured carbon samples have been investigated \cite{KBLLDZ[4]} - \cite{KBLLDZ[5]}  using a high-pressure microbalance. Electrochemical galvanostatic measurements at room temperature were done in \cite{KBLLDZ[6]} - \cite{KBLLDZ[9]}, where also  
volumetric (mass flow) gas phase measurements were done. It was shown that reversible physisorption takes place in all studied samples. The amount of adsorbed hydrogen is proportional to the surface area of the nanostructured carbon sample. The maximum specific surface area of carbon ($1315$ $m^{2} g^{-1}$ ) gives the maximum measured adsorption capacity of the nanostructured material: $2$ - mass percents. It was shown the great importance of various carbon nanostructures, including graphene, for hydrogen storage \cite{KBLLDZ[10]} - \cite{KBLLDZ[11]}, and \cite{FALG[11]}. Nevertheless the mechanism of hydrogen desorption is still unknown.

Recently (2018) E.A. Korzhnikova et al \cite{KBLLDZ} have studied clusters of discrete breathers in graphane. They used molecular dynamics simulation. The energy exchange between discrete breathers is studied for clusters composed of two and three discrete breathers. It is shown that difference in the initial amplitude or in the initial vibration phase of discrete breathers affect the energy exchange between them. The life time of single discrete breather in thermal equilibrium is studied. This study gives an example of a study of the discrete to continuum limit (one DB to many DB to continuum).

The Inverse Scattering Method we discussed in \cite{MJO1} (part I) (Methodological part with an example: Soliton solution of the Sine-Gordon Equation.), and  in \cite{MJO2}  (Methodological part II. with an example: Breather solution of the Sine-Gordon Equation). These kink and  breather solutions are also well known. We will use those papers in the continuum limit of a model describing graphane. We will also describe the phonon spectrum of graphane. Then we will discuss the relation of this limit description to the discrete limit. We will discuss physical properties of graphane as they are measured and with using our results.  In other to do this we need discuss also experimental properties of graphane as they are measured.

Initial structure of DBs in graphane can be found in \cite{KBLLDZ[15]}. DB can be excited by applying an out-of-plane displacement on a single hydrogen atom of graphane. The vibration frequency of the DB lies either within the gap of the phonon spectrum of graphane or beyond its upper spectrum bound. Both soft and hard types of anharmonicity of the DB are observed in graphane. The study shows that the DB has its lifetime which is affected by various factors including its anharmonicity type, its amplitude and frequency, and the force on the hydrogen atom that forms it. Competition of these factors results in a complex mechanism for the lifetime determination.

Graphene  was mechanically extracted from graphite \cite{FALG[2]}. Its physical properties are described in \cite{KBLLDZ[15][2]} - \cite{KBLLDZ[15][6]}, \cite{KBLLDZ[22]}, \cite{KBLLDZ[15][8]} - \cite{KBLLDZ[15][9]}. Graphene reacts with atomic hydrogen, which transforms this highly conductive zero-overlap semimetal into a semiconductor. In graphane, theoretically predicted in \cite{KBLLDZ[15][10]} and \cite{FALG[5]}, and experimentally found \cite{FALG[11]}, hybridization of carbon atoms leads to transformation from $sp^{2}$ to $sp^{3}$ on the $C$ atoms \cite{KBLLDZ[15][13]} - \cite{KBLLDZ[15][18]}. By means of hydrogenation of graphene, it is possible to alter its electronic transport \cite{KBLLDZ[15][19]}, tune its band gap \cite{KBLLDZ[15][20]}, change its electrochemical properties \cite{KBLLDZ[15][21]} - \cite{KBLLDZ[15][22]} and thermal conductivity \cite{KBLLDZ[15][15]} and \cite{KBLLDZ[15][18]}, and induce its room temperature ferromagnetism \cite{KBLLDZ[15][23]}. Hydrogenation of graphene can be achieved by experimental techniques such as exposure to cold hydrogen plasma, \cite{FALG[11]} and \cite{KBLLDZ[15][19]} and \cite{KBLLDZ[15][24]} - \cite{KBLLDZ[15][26]}, by using a conditioning catalyst upstream \cite{KBLLDZ[15][27]}, synchronized reduction and hydrogenation of graphene oxide in an aqueous suspension under gamma ray irradiation \cite{KBLLDZ[15][21]}, by applying an electric field as a catalyst \cite{KBLLDZ[15][28]} and by electron irradiation of an adsorbate on graphene \cite{KBLLDZ[15][14]}. Experimentally hydrogenation  was determined by the hydrogen coverage and morphology study of hydrogenated graphene  \cite{KBLLDZ[15][19]}.
Raman spectroscopy was used for estimating hydrogen coverage by measuring the relative intensities of bands, which are symmetry forbidden in pristine graphene and appear in the Raman spectrum of defected graphene due to the defect induced
symmetry breaking, in comparison with the G band that exists in pristine and hydrogenated graphene \cite{FALG[11]}. 
Theoretical studies on the structure and properties of graphene functionalized by means of hydrogenation have been done in \cite{KBLLDZ[15][10]} and \cite{KBLLDZ[15][18]} and  \cite{KBLLDZ[15][29]} - \cite{KBLLDZ[15][32]}. Particularly, the tendency for clustering of hydrogenated and nonhydrogenated sites was observed, \cite{KBLLDZ[15][13]} - \cite{KBLLDZ[15][30]}. This tendency is also interesting for study of the discrete to continuum limit (one DB to many).

Hydrogen storage technologies are cleaner and safer energy sources than others. Particularly, carbon materials, for example graphene, provide a playground for hydrogen storage due to their lightweight structure and high performance potential, \cite{FALG[11]} and \cite{KBLLDZ[15][33]} - \cite{KBLLDZ[15][37]}. It was shown that graphene at low temperatures
can easily absorb hydrogen and at high temperatures can easily desorb hydrogen \cite{FALG[11]}. Thus the study of hydrogenation and dehydrogenation of graphene is important. There is then interesting again  study of the discrete to continuus limit, from one DB to many to continuum. Dehydrogenation kinetics during annealing turns out to be not simple, the thermal stability
of adsorbed hydrogen can depend on the conditions of treatment \cite{KBLLDZ[15][26]}. It was found that there are two types of dehydrogenation mechanisms with different dehydrogenation barriers \cite{KBLLDZ[15][25]}. Then it would be interesting to find a theoretical explanation of this effect. For vacuum annealing dehydrogenation starts at low temperatures \cite{KBLLDZ[15][25]}. To achieve complete dehydrogenation within an annealing time on the order of $1$ $h$, the annealing temperature
should be increased up to $400 - 500$ $^{o}C$ , \cite{KBLLDZ[15][25]} and \cite{KBLLDZ[15][26]}. The annealing temperature cannot be raised higher than $500$ $^{o}C$ because of damage to graphene sheets by thermal
fluctuation and by interaction with their substrate increases with temperature, \cite{FALG[11]} and \cite{KBLLDZ[15][26]}. Heating in nitrogen environment reduces the annealing time to minutes \cite{KBLLDZ[15][19]}; this all is still beyond the capacity of molecular dynamics simulation. Thus it is necessary to study these systems and mechanisms also by other methods.

The hydrogen atoms, they are $12$ times lighter than the carbon atoms C, 
a this large difference in their atomic weights leads to wide gap in the phonon band, and can be expected and is a necessary
condition for the existence of nonlinear excitations called gap
discrete breathers (DBs) or intrinsic localized modes, \cite{KBLLDZ[15][38]} and \cite{KBLLDZ[15][39]}, which are able to localize a large amount of energy so as to probably play an important role in overcoming the potential
barrier of hydrogen desorption. DBs are spatially localized vibrational modes of large amplitude in nonlinear defect-free lattices, they were described theoretically for the first time in \cite{KBLLDZ[15][40]}. The DBs were discussed in  many physical systems, \cite{KBLLDZ[15][39]} - \cite{KBLLDZ[15][41]}. The existence of stationary intrinsic localized modes has been proved experimentally in thermodynamic equilibrium in NaI crystals \cite{KBLLDZ[15][42]}. According to computer simulations, DBs can also be excited in graphene, \cite{KBLLDZ[19]} and \cite{KBLLDZ[15][44]}. Linear and nonlinear vibrational modes can exist at the edges of graphene nanoribbons, \cite{KBLLDZ[15][45]} - \cite{KBLLDZ[15][46]}, and in other carbon nano-polymorphs, \cite{KBLLDZ[24]} and \cite{KBLLDZ[25]}. 
In graphene, there is no gap in the phonon spectrum, there is a gap in graphane because of the presence of
$H$ atoms \cite{KBLLDZ[15][49]}.

It is observed \cite{KBLLDZ} that in order to form a stable DB, the maximum distance $r$ between atoms $C$ and $H$ reach 1.62 $\AA{}$, at which the amplitude $A$ reaches the maximum value of 0.40 $\AA{}$. The attempt to increase $r$ beyond 
$r = 1.62$ $\AA{}$ by applying a larger D0 would cause the atom $H$ to debond from $C$ atom (and from the whole sheet). Debonding mechanism is important for hydrogen desorption. At $r = 1.62$ $\AA{}$, the total energy of the graphane system is found to increase by about 3.85 $eV$, compared with that of the system in which no DB is excited. This implies that the atom $H$ can be debonded (can escape from C) if the provided energy is larger than $3.85$ $eV$. DBs can be spontaneously excited in crystals at finite temperatures. This will  localize a large amount of energy and the hydrogen atom can overcome the energy barrier of desorption from carbon $C$ atoms. High temperatures thus may activate or accelerate dehydrogenation of hydrogenated graphene at high temperatures.

DBs are excitations of defect-free lattices. In fact for localized to one atom DB the lattice is not necessary. Evidence for hydrogen clustering at relatively high hydrogen coverage was found experimentally \cite{KBLLDZ[15][22]}. A question arises as to whether DBs partially dehydrogenate graphane. The study \cite{KBLLDZ} shows that DBs in graphane are highly localized so that even relatively small hydrogenated islands on graphene can host DBs. There are two activation energies of dehydrogenation with the transition temperature at about $200^{o}$ $C$ \cite{KBLLDZ[15][25]}. The smaller dehydrogenation activation energy at temperatures below $200^{o}$ $C$ observed in \cite{KBLLDZ[15][19]} and \cite{KBLLDZ[15][25]} can be understood by the metastable attachment of the energetic ions to the graphene sheet during plasma hydrogenation. However, the larger dehydrogenation activation energy at temperatures above $200^{o}$ $C$ has not yet been well explained. Note that an atom can be absorbed on the surface (chemical process) or adsorbed on the surface (physical process). DBs can be spontaneously excited at a finite-temperature thermal equilibrium \cite{KBLLDZ[15][57]} - \cite{KBLLDZ[15][60]}. The maximum kinetic energy of the externally excited DBs is found \cite{KBLLDZ} to be very close to the energy barrier for hydrogen desorption. Thus spontaneously excited DBs in graphane at finite temperatures may play a role in activating or accelerating the dehydrogenation of hydrogenated graphene. Mechanical properties of graphene were studied in \cite{BLDZ}. In \cite{WY} electronic structures of clusters of hydrogen vacancies on graphene were studied.

Graphane is the end product of the complete hydrogenation of graphene \cite{FALG[11]}, \cite{KBLLDZ[15][24]}, and \cite{WY[3]} - \cite{WY[6]}, \cite{FALG[5]}, \cite{WY[8]} -\cite{WY[20]}. Carbon atoms in graphane are bonded to $H$ atoms alternately from either side of the plane of graphene. Metal-insulator transition \cite{FALG[11]}, \cite{KBLLDZ[15][24]}, \cite{WY[3]} and \cite{WY[4]}, occurs as a result of the process, opening a large band gap \cite{WY[6]}, \cite{FALG[5]}, cite{WY[8]} - \cite{WY[11]} for graphane. Some $H$ atoms can desorb from one side of graphane as a result of an applied electric field \cite{FALG[17]},\cite{WY[13]}, \cite{WY[20]}, and the desorption process may continue to the extent. Patterns of $H$ vacancies in graphane were studied in \cite{FALG[17]}, \cite{WY[13]} -\cite{WY[15]}, \cite{WY[19]}, \cite{WY[21]} - \cite{WY[23]}, \cite{KBLLDZ[15][14]} and \cite{WY[3]}, \cite{WY[6]}, \cite{WY[10]} - \cite{WY[32]} as graphane with patches or clusters of $H$ vacancies. The results can be applied to the design of nanoelectronic circuits \cite{WY[9]} - \cite{WY[10]}, \cite{WY[25]} and may serve as a guide for predicting properties of larger and more complicated patterns of the $C-H$ composites. $H$ vacancies can be continuously distributed over a vast area and simulated by a large periodic structure. $H$ vacancies can also be confined in a finite area of graphane. Continuous presence of $H$ vacancies can tune the width of band gap of graphane \cite{WY[33]}, \cite{WY[34]} and, in the case of a single H-vacancy chain, even turns the defected graphane into a conductor with linear band dispersion \cite{WY[33]} near the Fermi level.

In \cite{SK} authors model, with the use of the force field method, the dependence of mechanical conformations of graphene sheets, located on flat substrates, on the density of unilateral (one-side) attachment of hydrogen, fluorine or chlorine atoms to them. There are various derivatives of graphene (hexagonal monolayer of carbon atoms) \cite{FALG[2]}, \cite{FALG[3]}, \cite{KBLLDZ[15][4]} and \cite{STN[51]}, such as graphane $CH$ and fluorographene $CF$ (a monolayer of graphene, saturated on both sides with hydrogen or fluorine) \cite{FALG[5]}, \cite{FALG[11]}, \cite{SK[7]} - \cite{SK[8]}, graphone $C_{2}H$ (graphene monolayer saturated with hydrogen on one side) \cite{WY[13]}, \cite{WY[10]}, \cite{SK[11]} - \cite{SK[12]}, one-side fluorinated graphene $C_{4}F$ \cite{SK[13]} and \cite{SK[14]}, chlorinated graphene $C_{4}Cl$ \cite{SK[15]}. Valence attachment of an external atom to a graphene sheet leads to the local convexity of the sheet as result of the appearance of the $sp^{3}$ hybridization at the joining point \cite{SK[16]} and \cite{KBLLDZ[15][32]}. Therefore, if hydrogen atoms are attached on one side in the finite domain of the sheet, creating a local peace of graphone on the sheet, a characteristic convex deformation of the sheet occurs in this region \cite{SK[18]}. If the hydrogen, fluorine or chlorine atoms are attached uniformly to one whole side of the sheet, the whole small sheet will take a convex shape while a large sheet will fold into a roll \cite{SK[19]} - \cite{SK[22]}.

Authors of \cite{SK} use the force field AMBER (Assisted Model Building with Energy Refinement) \cite{SK[41]} and \cite{SK[42]}. In this model, the strain energy of the valence $sp^{2}$ and $sp^{3}$ $C-C$ and $C-CR$ bonds, and of $O-H$, $C-R$ bonds (here an atom or group of atoms $R = H, F$) is described by the Morse potential. The nonvalent van der Waals interactions of atoms are described by the Lennard-Jones potential.
The values of the potential parameters for carbon atoms of the graphene layer authors take from\cite{SK[45]}, for the remaining atoms are taken from \cite{SK[46]}.

In real crystals there are atoms adsorbed (physically) on a crystal and absorbed (chemically) on its surface. The first case corresponds to heterostructures with the Lennard-Jones potential and the second case corresponds to the Morse potential. Note however that the potential and thus the force ($C-H$ in graphane) which was found in \cite{KBLLDZ} is not the Lennard-Jones potential nor the Morse potential. Then it is necessary to model the potential and thus the force ($C-H$ in graphane).
In \cite{KMacK} authors find conditions for existence and stability of various types of Discrete Breather concentrated around three central sites in a triangular lattice of one- dimensional Hamiltonian oscillators with on-site potential and nearest-neighbour coupling. Authors \cite{DPW} study localized breather-like solution in a discrete Klein-Gordon model.

In this paper we will study, using both our papers mentioned above, localized and nonlocalized breather-like solutions in a discrete and continuous Frenkel-Kontorova model for graphane. We will use the potential of $H$ atom above the $C$ atom in the graphane $C$ lattice which model calculation results of \cite{KBLLDZ}. The potential found by \cite{KBLLDZ} does not have the Morse expression-like nor the Lennard-Jones potential expression-like form. Using \cite{DPW} we will consider model and excitations of the system (graphane) in the next section. Study of the large amplitude solution for DB  is in the next section. Dynamics of a kink and breather in the continuum limit will be done in the next sections. Solution of the discrete breather will be formulated in the next section. Application of results to graphane will be done in the next section. Summary and Conclusions of the paper are done in the last section.

\section{Graphane: Hexagonal Lattice, Lagrangian.}

The graphane has a hexagonal lattice of atoms C. We neglect deformations of the graphene lattice due to presence of $H$ atoms, as they were discussed in Introduction. Let us denote the position of the $H$ atom above the $C$ atom in the  direction perpendicular to the hexagonal lattice of $C$ atoms as $x({\bf r}_{i,j})$ for $C$ atom at the site of the lattice ${\bf r}_{i,j}$. Here $(i, j)$ are integers, $( i, j) \in Z^{2}$. This $H$ atom interacts with neighbouring $H$ atoms. Let us denote the coupling constant of  this interaction as $\epsilon$. The hexagomal lattice will be described later.

The potential energy of the $H$ atom at the site ${\bf r}_{i,j}$  of the lattice is $V(x({\bf r}_{i,j}))$. Using the results of the calculation \cite{KBLLDZ} we assume that the potential energy possesses a stable minimum at the point $x({\bf r}_{i,j}) = x_{0}$. Also we assume using these results that $\frac{d V(x({\bf r}_{i,j})}{d x({\bf r}_{i,j})} > 0$ for $x({\bf r}_{i,j}) > x_{0}$, and that $\frac{d V(x({\bf r}_{i,j})}{d x({\bf r}_{i,j})} < 0$ for $x({\bf r}_{i,j}) < x_{0}$. Further we assume that $\frac{d^{2} V(x({\bf r}_{i,j})}{d x({\bf r}_{i,j})^{2}} \equiv \omega_{ph}^{2} > 0$ for $x({\bf r}_{i,j}) = x_{0} $. We will assume that the results of \cite{KBLLDZ} may be described by the potential $V(x({\bf r}_{i,j}))\approx V_{0} + V_{1}(1-
\cos(\frac{x({\bf r}_{i,j})-x_{0})}{\lambda}))$.
Here $V_{0}, V_{1},\lambda$ and $x_{0}$ are parameters of the potential energy. The first and the second parameter describe the potential scale, the third parameter is the width of the 
change of the $x({\bf r}_{i,j}) $ variable (which is distance of the atom $H$ from the atom $C$ below). The fourth parameter is the parameter of the minimum of the potential.
In \cite{KBLLDZ} it was calculated that the energy $E(r)$ is nonzero for the range $r_{0}< r< r_{1}$. We will use the interval of $x({\bf r}_{i,j}-x_{0}) $ variation as $- \frac{\pi}{2} \lambda < x({\bf r}_{i,j}-x_{0}) < \beta . \frac{\pi}{2} \lambda$. Here the parameter $0 < \beta < 1$.

The momentum of the $H$ atom motion in the direction perpendicular to the $2d$ hexagonal plane at the site ${\bf r}_{i,j}$ is $ p({\bf r}_{i,j}) $. The Hamiltonian $H_{H}$ of the system of $H$ atoms (of atoms $H$ on the graphane one side, we do not consider the interactions of $H$ atoms of the side-side type) is:
\begin{equation}\label{1}
H_{H} = \sum_{ij}^{+ \infty} [ \frac{p({\bf r}_{i,j})}{2} + 
V(x({\bf r}_{i,j}))] + \frac{\epsilon}{2} \sum^{+\infty}_{ij} [ (x({\bf r}_{i,j}) - x({\bf r}_{i+1,j}))^{2} +
(x({\bf r}_{i,j}) - x({\bf r}_{i-1,j}))^{2}
\end{equation}
\[  + (x({\bf r}_{i,j}) - x({\bf r}_{i,j+1}))^{2}]. \]
The sum $\sum_{ij}$ is over the vertices $( i, j) \in Z^{2}$ of the hexagonal lattice.

Let us describe the $2d$ hexagonal lattice as in \cite{DFP}. 
A graphene sheet is a discrete set of carbon atoms $C$ that, in the absence of external forces, are at the vertices of a periodic array of hexagonal cells. Atoms occupy the nodes of the $2d$ lattice,  generated by two simple Bravais lattices $L_{1} $ and $L_{2} $: 
\begin{equation}\label{2}
L_{1} (l) = {{\bf r}_{ij} \in R^{2} :{\bf r} = i. l. {\bf d_{1}} +
 j. l. {\bf d_{2}} }
\end{equation}

with $( i, j) \in Z^{2}$, and
\begin{equation}\label{3}
L_{2} (l) = l. {\bf p} + L_{1} (l),
\end{equation}

here $l$ is a lattice constant by which both Bravais lattices are
shifted one with respect to another one. In (\ref{2}) ${\bf d}_{1}$, ${\bf d}_{2}$ and ${\bf p}$ respectively are the lattice vectors and the shift vector of the two dimensional (2d) hexagonal lattice, whose Cartesian components are given by ${\bf d}_{1}=(\sqrt{3},0)$, ${\bf d}_{2}=(\frac{\sqrt{3}}{2},\frac{3}{2})$  and 
 ${\bf p}=(\frac{\sqrt{3}}{2},\frac{1}{2})$.

 The sides of the hexagonal cells in the lattice stand for the bonds between pairs of the $C$ atom and the neighbor $C$ atoms, and are represented by  vectors \cite{DFP}: 
\begin{equation}\label{4}
{\bf p_{\alpha}}= {\bf d_{\alpha}} - {\bf p}
\end{equation} 
where $(\alpha = 1, 2)$, and ${\bf p_{3}} = - {\bf p}$, and
\begin{equation}\label{4,}
{\bf d_{3}}= {\bf d_{2}} - {\bf d_{1}}.
\end{equation} 

As a reference configuration the set of points ${\bf x}^{l}$ $ \in $ 
$ L^{1}(l) \cup L^{2}(l) $ is taken.

The Lagrangian $L_{H}$ of the system is:
\begin{equation}
\label{5}
L_{H} = \sum^{+\infty}_{ij} [ \frac{m}{2} (\frac{\partial x({\bf r}_{i,j})}{\partial t })^{2} - V(x({\bf r}_{i,j})) ] - \frac{\epsilon}{2} \sum^{\infty}_{ij} [ (x({\bf r}_{i,j}) - x({\bf r}_{i+1,j}))^{2} 
\end{equation}
\[ + (x({\bf r}_{i,j}) - x({\bf r}_{i-1,j}))^{2} + (x({\bf r}_{i,j}) - x({\bf r}_{i,j+1}))^{2}].  \]
Here $t$ is time, $m$ is the effective mass of the atom H.

The potential energy $V(x({\bf r}_{i,j}))$ of the $H$ atom at the site of the lattice ${\bf r}_{i,j}$ may be approximated as described above:
\begin{equation}\label{7}
V(x({\bf r}_{i,j}()) = V_{0} + V_{1}(1-\cos(\frac{x({\bf r}_{i,j})-x_{0})}{\lambda})).
\end{equation}

For the continuous limit of the Lagrangian in the next section we will use the expansion of vectors ${\bf r}_{i \pm 1,j}, {\bf r}_{i ,j + 1}$ in the lattice parameter $l$: \[{\bf r}_{i + 1,j} \approx {\bf r}_{i,j} + l.{\bf d_{1}}, \]
\[ {\bf r}_{i - 1,j} \approx {\bf r}_{i,j} - l.{\bf d_{1}} , \]
and
\[{\bf r}_{i,j+1} \approx {\bf r}_{i,j} + l.{\bf d_{2}} . \]

\section{Frenkel-Kontorova Model and Graphane.}

The Lagrange-Euler equations are found from (\ref{5}). They have the form (we do not take into account dissipative forces):
\begin{equation}
\label{8}
m \frac{\partial^{2} x_{ij}}{\partial t ^{2}} + \frac{d V(x_{ij})}{x_{ij}} + 
\epsilon [ (x_{ij} - x_{i+1 j}) + (x_{ij} - x_{i-1 j}) + (x_{ij} - x_{i j+1})] =0.
\end{equation}

After substitution of the potential  energy $V(x({\bf r}_{i,j}))$ from (\ref{7}) we obtain the Lagrange-Euler equations in the form:
\begin{equation}
\label{9}
m \frac{\partial^{2} x_{ij}}{\partial t ^{2}} + \frac{V_{1}}{\lambda} \sin(\frac{x({\bf r}_{i,j}) - x_{0}}{\lambda}) + 
\epsilon [ (x_{ij} - x_{i+1 j}) + (x_{ij} - x_{i-1 j}) + (x_{ij} - x_{i j+1})] =0.
\end{equation}

For small lattice constant approximation (the end of the previous section) we obtain the Sine-Gordon equation (we denote $x({\bf r}_{i,j}) = x({\bf r})$ ), where the vector ${\bf r}= (x,y)$ has $x, y$ coordinates in the plane, they are orthogonal: 
\begin{equation}\label{10}
m \frac{\partial^{2} x_{ij}}{\partial t ^{2}} + \frac{V_{1}}{\lambda} \sin(\frac{x({\bf r}) - x_{0}}{\lambda}) -
\epsilon \frac{27 l^{2}}{2} [\frac{\partial^{2} x({\bf r}) }{\partial x ^{2}} + \frac{\partial^{2} x({\bf r}) }{\partial y ^{2}} + \frac{1}{3 \sqrt{3}} \frac{\partial x({\bf r}) }{\partial x} \frac{\partial x({\bf r}) }{\partial y}]  =0.
\end{equation}

\section{The One $H$ Physical Motion.}

  Let us study the equations of motion (\ref{10}) for the case of one $H$ atom (we denote its coordinate as $x({\bf r}_{i,j}) = x_{r} $ ), where ${\bf r}= (x,y)$ with $x, y$ coordinates in the plane. The Lagrange-Euler equation is found from (\ref{9}) in the form: : 
  \begin{equation}\label{11}
 m \frac{d^{2} x_{r}}{d t ^{2}} +  \frac{V_{1}}{\lambda} \sin(\frac{x_{r}-x_{0}}{\lambda})   =0.
  \end{equation}

Let us introduce the variable $y_{r} = \frac{x_{r}-x_{0}}{\lambda}$. Then the equation (\ref{11}) will take the form:
  \begin{equation}\label{12}
m \frac{d^{2} y_{r}}{d t ^{2}} +  \frac{V_{1}}{\lambda^{2}} \sin(y_{r})   =0.
\end{equation}

From (\ref{12}) we obtain by multiplying with $ \frac{d y_{r}}{d t } $ the equation:
\begin{equation}\label{13}
\frac{d y_{r}}{d t } = \pm \sqrt{\frac{2 V_{1}}{m \lambda^{2}}(D - \cos(y))}
\end{equation}

where $D$ is a constant. We will assume that $D > 1$.

The integral $I$ which we obtain from (\ref{13}) is:
\begin{equation}\label{14}
I = \int_{y_{0}}^{y_{h}} \frac{d y}{\sqrt{D - \cos(y)}} = \pm \sqrt{\frac{2 V_{1}}{m \lambda^{2}}} (t-t_{0}).
\end{equation}

Here $t_{0}$ is a time constant, \[y_{0} = \frac{x(t_{0})-x_{0}}{\lambda}\] and \[y_{h} = \frac{x(t)-x_{0}}{\lambda}.\] The $k$ parameter will be defined as \[k^{2} = \frac{2}{D - 1}.\] The integral $I$ as calculated has the form:
\begin{equation}
\label{15}
I = \frac{2}{\sqrt{D-1}} (F(\frac{y_{h}}{2},k)) - F(\frac{y_{0}}{2},k)) = \pm \sqrt{\frac{2 V_{1}}{m \lambda^{2}}} (t - t_{0}).
\end{equation}

Here $F(\frac{y_{h}}{2},k)$ and $F(\frac{y_{h}}{2},k)$  are first elliptic integrals.
Note that we assume here $0 < k^{2} < 1$. Let us introduce variables $u_{k}$ and $u_{0}$,  $\sin(\frac{y_{0}}{2}) = sn(u_{0})$ and $\sin(\frac{y_{h}}{2}) = sn(u_{h})$, where $sn$ function is $\sin$ elliptic function.
From (\ref{15}) we obtain:
\begin{equation}\label{16}
u_{h} - u_{0} = \pm \frac{\sqrt{D - 1}}{2} \sqrt{\frac{2 V_{1}}{m \lambda^{2}}} (t - t_{0}).
\end{equation}

From (\ref{16}) we will discuss the case $1$, when the $H$ atom is at the position $x(t_{0})$ for which it holds $x_{0} < x(t_{0}) < \beta \frac{\pi}{2} $, and
the case $2$, when the $H$ atom is at the position $x(t_{0})$ for which it holds $x_{0} > x(t_{0}) > - \frac{\pi}{2} $, and the case $3$, when the $H$ atom is at the position $x(t_{0})$ for which it holds the it is small quantity near 
$ x(t_{0}) \approx x_{0}$.

The case $1$, when the $H$ atom is at the position $x(t_{0})$ for which it holds $x_{0} < x(t_{0}) < \beta \frac{\pi}{2} $ corresponds to an $H$ atom which is above the position $x_{0}$ but below the position at which it is desorbed from the graphane. Then the motion of the $H$ atom at time $t_{h}$ is described by $x(t_{h})$ from $y(t_{h}) = \frac{x(t_{h}) - x_{0}}{\lambda}$ where $\sin(\frac{y_{h}}{2}) = sn(u_{h})$ is given from (\ref{16}).

The case $2$, when the $H$ atom is at the position $x(t_{0})$ for which it holds $x_{1} < x(t_{0}) < - \frac{\pi}{2} $ corresponds to an $H$ atom which is above the position $x_{1}$. This position is a position of the suppressed atom $H$ in the direction to the $C$ (nearer) atom. This $H$ atom will be desorbed from the graphane going through the minimum $x_{0}$ of the potential energy $V(x({\bf r}_{i,j}))$ and then through the point $\beta \frac{\pi}{2}$ (the point of desorption) of the potential energy $V(x({\bf r}_{i,j}))$. The motion of the $H$ atom at time $t_{h}$ is described by $x(t_{h})$ from $y(t_{h}) = \frac{x(t_{h}) - x_{0}}{\lambda}$ where $\sin(\frac{y_{h}}{2} = sn(u_{h}))$ is given from (\ref{16}).

The case $3$, when the $H$ atom is near the position $x_{0}$ at the time $t_{0}$ corresponds to an $H$ atom which is oscillating near this position in graphane. The motion of the $H$ atom at time $t_{h}$ is described by $x(t_{h})$ from $y(t_{h}) = \frac{x(t_{h}) - x_{0}}{\lambda}$ where $\sin(\frac{y_{h}}{2}) = sn(u_{h})$ is given from (\ref{16}). 

\section{Continuous limit of the  Lagrange-Euler equations.}

From (\ref{9}) we obtain using small lattice constant approximation the Sine-Gordon equation (we denote $x({\bf r}_{i,j}) = x({\bf r})$ ), where ${\bf r}= (x,y)$ with $x, y$ coordinates in the plane, the equation (\ref{10}). There is a term 
$\frac{1}{3 \sqrt{3}} \frac{\partial x({\bf r}) }{\partial x} \frac{\partial x({\bf r}) }{\partial y}$ in this equation. This term is a non-diagonal term in the $(x, y)$ coordinates. We transform the coordinates $x, y$ to such coordinates that this non-diagonal term will not be present in new coordinates equation (\ref{10}). New coordinates we define as $(c, d)$ and define as:
\begin{equation}\label{17}
x = \eta c + \theta d
\end{equation} 

and \[y =\eta c - \theta d. \]
Here we take the constants $\eta$ and $\theta$ from equations $\eta^{2} (2+\frac{2}{3 \sqrt{3}})=1 $ and $\theta^{2} (2 - \frac{2}{3 \sqrt{3}})=1$. The Sine-Gordon equation (\ref{10}) has in coordinates $(c, d)$ the form (here we introduce a new function $f(c, d)$ putting $x({\bf r}_{i,j}) \equiv f(c, d) + x_{0}$):
\begin{equation}\label{18}
m \frac{\partial^{2} f(c, d)}{\partial t ^{2}} + \frac{V_{1}}{\lambda} \sin(\frac{f(c, d)}{\lambda}) + 
\epsilon \frac{27 l^{2}}{2} [\frac{\partial^{2} f(c, d) }{\partial c ^{2}} + \frac{\partial^{2} f(c, d) }{\partial d ^{2}} ]  =0.
\end{equation}
As we can see the non-diagonal term is absent in (\ref{18}).
Denoting $c_{1} \equiv \frac{V_{1}}{m \lambda}$, $f(c, d) \equiv F(X, Y)$, taking new variables $(X, Y)$ by $X = \frac{c}{\sqrt{\epsilon \frac{27 l^{2}}{2m}}}$ and by $Y = \frac{d}{\sqrt{\epsilon \frac{27 l^{2}}{2m}}}$ we obtain the Sine-Gordon equation (\ref{10}) in the form:
\begin{equation}\label{19}
\frac{\partial^{2} F(X, Y)}{\partial t ^{2}} + c_{1}\sin(\frac{F(X, Y)}{\lambda}) - 
\frac{\partial^{2} F(X, Y) }{\partial X ^{2}} - \frac{\partial^{2} F(X, Y)}{\partial Y ^{2}}   =0.
\end{equation}

\section{Sine-Gordon Equation and Its Soliton Solutions.}

Recently we discussed \cite{MJO1} the Inverse Scattering Method (part I) (Methodological part with an example: Soliton solution of the Sine-Gordon Equation.).
Let us introduced rescaled variables $u(X^{'}, Y^{'}) = \frac{F(X, T)}{\lambda}$, $X^{'} = X \sqrt{c_{1}}$, $Y^{'} = Y \sqrt{c_{1}}$ and $T = t \sqrt{c_{1}}$.
The Sine-Gordon equation (\ref{19}) has now the form:
\begin{equation}\label{20}
\frac{\partial^{2} u(X^{'}, Y^{'})}{\partial t ^{2}} + \sin(u(X^{'}, Y^{'})) - 
\frac{\partial^{2} u(X^{'}, Y^{'}) }{\partial X^{'2}} - \frac{\partial^{2} u(X^{'}, Y^{'})}{\partial Y^{'2}}   =0.
\end{equation}

The kink solution of the Sine-Gordon equation (\ref{20}) in $X^{'}$ direction is \cite{MJO1}:
\begin{equation}\label{21}
u(X^{'}) = 4 \arctan (\exp(\frac{(X^{'}-X_{0}) - U.(T-T_{0})}{\sqrt{1 - U^{2}}})).
\end{equation}

Here $U$ is a velocity of the kink, $ U^{2} < 1$, $T_{0}$ and $X^{'}_{0}$ are in time $T$ and in $X^{'}$ direction constants of the kink.  A similar solution describes a kink in time $T$ and in $Y^{'}$ direction.

\section{Sine-Gordon Equation and Its Breather Solution.}

Recently we discussed \cite{MJO2} the Inverse Scattering Method (part II) (Methodological part with an example: Breather solution of the Sine-Gordon Equation). While (\ref{16}) describes a DB, the breather solution in the continuum limit is described in the following.

We will use introduced above rescaled variables $u(X^{'}, Y^{'}) = \frac{F(X, T)}{\lambda}$, $X^{'} = X \sqrt{c_{1}}$, $Y^{'} = Y \sqrt{c_{1}}$ and $T = t \sqrt{c_{1}}$.
The Sine-Gordon equation (\ref{19}) has the breather solution
in $X^{'}$ direction, \cite{MJO2}:
\begin{equation}\label{21}
u(X^{'}) = 4 \arctan (\exp(\frac{(\sqrt{\frac{1-\omega^{2}}{\omega^{2}}}.\cos(T-T_{0}))}{\cosh(\sqrt{1 - \omega^{2}}.(X^{'} - X^{'}_{0}))})).
\end{equation}

Here $\omega$ is the frequency of the breather, $T_{0}$ and $X^{'}_{0}$ are in time $T$ and in $X^{'}$ direction constants of the breather.

\section{ Summary and Conclusions.}

In this paper we studied the graphane - material for hydrogen storage. The Frenkel-Kontorova model on hexagonal lattice was used. Due to discussed in the Introduction continuous and discrete limits in the $2d$ graphane lattice we studied the one $H$ physical motion which is the case of one $H$ atom above the $C$ atom in the plane of graphane (we used the approximation of the hexagonal lattice in the plane). Continuous limit of the  Lagrange-Euler equations found from the Hamiltonian for $H$ atoms motion enabled us to study kink and breather in the $H$ plane above the $C$ plane.

We have found that there are three cases in the one $H$ atom motion  The case $1$, when the $H$ atom is at the position $x(t_{0})$ for which it holds $x_{0} < x(t_{0}) < \beta \frac{\pi}{2} $ (and similarly when $x_{0} < x(t_{0}) < - \beta \frac{\pi}{2} \lambda $) corresponds to an $H$ atom which is above the position $x_{0}$ but below the position at which it is desorbed (at the point $\beta \frac{\pi}{2} \lambda$) from the graphane. Then the motion of the $H$ atom at time $t_{h}$ is described by $x(t_{h})$ from $y(t_{h}) = \frac{x(t_{h}) - x_{0}}{\lambda}$ where $\sin(\frac{y_{h}}{2}) = sn(u_{h})$ which satisfies the condition $ -\beta \frac{\pi}{2} \lambda < x(t_{h}) < \beta \frac{\pi}{2} \lambda$. The case $2$, when the $H$ atom is at the position $x(t_{0})$ for which it holds $x_{1} < x(t_{0}) < - \frac{\pi}{2} $ which corresponds to an $H$ atom which is above the position $x_{1}$. This position is a position of the suppressed atom $H$ in the direction to the $C$ (nearer) atom. This $H$ atom will be desorbed from the graphane going through the minimum $x_{0}$ of the potential energy $V(x({\bf r}_{i,j}))$ and then through the point $\beta \frac{\pi}{2}$ (the point of desorption) of the potential energy $V(x({\bf r}_{i,j}))$. The motion of the $H$ atom at time $t_{h}$ is described by $x(t_{h})$ from $y(t_{h}) = \frac{x(t_{h}) - x_{0}}{\lambda}$ where $\sin(\frac{y_{h}}{2} = sn(u_{h}))$ is given from (\ref{16}). The case $3$, when the $H$ atom is near the position $x_{0}$ at the time $t_{0}$ corresponds to an $H$ atom which is oscillating near this position in graphane. The motion of the $H$ atom at time $t_{h}$ is described by $x(t_{h})$ from $y(t_{h}) = \frac{x(t_{h}) - x_{0}}{\lambda}$ where $\sin(\frac{y_{h}}{2}) = sn(u_{h})$ and is given from (\ref{16}). The position of the atom $H$ at the time $t_{0}$ is the position to which the atom $H$ was excited with external force. While the results of the case $1$, and $3$, describe the motion of the $H$ atom at the graphane, the result $2$, described a new type of desorption process.
The lattice of $H$ atoms may be excited as described by the kink solution of the Sine-Gordon equation (\ref{20}) in $X^{'}$ direction as $u(X^{'}, T)$ from (\ref{21}). The kink has its velocity $U$, $ U^{2} < 1$, and the $T_{0}$ and $X^{'}_{0}$ in time $T$ and in $X^{'}$ coordinate direction localization.  A similar solution describes a kink in time $T$ and in $Y^{'}$ direction localized. This excitation fulfills the condition  $ -\beta \frac{\pi}{2} \lambda < x(t_{h}) < \beta \frac{\pi}{2} \lambda$. The Sine-Gordon equation (\ref{19}) has the breather solution in $X^{'}$ direction, $u(X^{'}, T)$ from (\ref{21}) where 
$\omega$ is the frequency of the breather, $T_{0}$ and $X^{'}_{0}$ are in time $T$ and in $X^{'}$ direction constants of the breather. This breather excitation fulfills the condition  $ -\beta \frac{\pi}{2} \lambda < x(t_{h}) < \beta \frac{\pi}{2} \lambda$.

\end{document}